\documentclass[12pt,letterpaper,makeidx]{article}
\usepackage[latin1]{inputenc}
\usepackage{amsmath}
\usepackage{amsfonts}
\usepackage{amssymb}
\usepackage{epsfig}
\usepackage{theorem}
\usepackage{oldgerm}
\usepackage{tabularx}
\usepackage{placeins}
\usepackage{amscd}
\usepackage{bm}
\usepackage{epic}
\usepackage{curves}
\usepackage{graphicx}
\usepackage{makeidx}
\usepackage{fancyhdr}
\usepackage{graphpap}
\usepackage{rotating}
\usepackage{bbm}
\usepackage{pstricks}
\usepackage[absolute]{textpos}
\usepackage{psfrag}
\usepackage{times}
\usepackage{dcolumn}
\usepackage{shadow}
\usepackage{multicol}
\usepackage[latin1]{inputenc}
\usepackage{oldgerm}
\fancypagestyle{plain}{%
\fancyhead{} 
}
\title{On the non-relativistic limit of charge conjugation in QED 
}  
\author{B. Carballo Pérez\footnote{brendacp@nucleares.unam.mx}  and M. Socolovsky\footnote{socolovs@nucleares.unam.mx} \\ 
\small{ Instituto de Ciencias Nucleares, Universidad Nacional Autónoma de México,}\\
\small{ Circuito exterior, Ciudad Universitaria, 04510, México D.F., México}}
\date{}

\begin{document}
\maketitle
\textbf{Abstract}\\
\hspace{0.5cm} Even if at the level of the non-relativistic limit of full QED, $C$ is not a symmetry, the limit of this operation does exist for the particular case when the electromagnetic field is considered a classical external object coupled to the Dirac field. This result extends the one obtained when fermions are described by the Schr\"odinger-Pauli equation. We give the expressions for both the $C$ matrix and the $\hat{C}$ operator for galilean electrons and positrons interacting with the external electromagnetic field. The result is relevant in relation to recent experiments with antihydrogen.\\


\newpage

\section{Introduction}

\hspace{0.5cm} It is very common to talk about P (parity or space inversion) and T (time inversion) symmetries, not only in the  non-relativistic quantum mechanics, but also in classical mechanics. Nevertheless, the existence of the galilean limit of C (charge conjugation) is nowadays controversial (\cite{Landau} vs \cite{CP violation}). Despite the fact that some authors \cite{Landau} consider that the C transformation does not have an equivalent in the non-relativistic theory, others \cite{CP violation} reach up to define the $\hat{C}$ operator in non-relativistic quantum mechanics, but they do not give an explicit expression for this operator. It is generally believed that the antiparticle concept can only be defined in the context of relativistic quantum mechanics, because it is only in this regime that we obtain solutions for free particles with negative energy flying backward in time, whose absence is interpreted as particles of opposite moment, charge and energy, flying forward in time: antiparticles. As the charge conjugation operation makes the particle-antiparticle operation, is believed that this operator only takes place in this context.

What is clear is that the C invariance in relativistic theory should not necessarily remain in the galilean limit because the transition from Lorentz group to Galilean group is not continuous.

The existence of the charge conjugation operation in the non-relativistic limit was demonstrated in 2006 \cite{L1, L2}, i.e., in the context of galilean relativity. This fact is different from the idea that C exists only in the context of relativistic physics, with the Lorentz-Poincaré group as the symmetry of space-time.

The Dirac equation interacting with a classical external electromagnetic field was considered in \cite{L1}, as the first step for finding this limit. For this purpose, the authors took into account the following facts:

\begin{itemize}
\item  $C$  does not belong to the Lorentz group $O(3,1)$.
\item We can give \emph{ad hoc} definitions for $C$, not only in the non-relativistic quantum mechanics, but also in the classical lorentzian and galilean mechanics \cite{CP violation}.
\item There is a symmetry between particles and antiparticles in the quantum relativistic theory. For the equations of both particles and antiparticles, it is possible to find a non-relativistic well defined limit, without any contradiction between them. If we have electrons of low energies (slow electrons), we can expect to have positrons of low energies (slow positrons) \cite{poslentos}, related by the charge conjugation operation.
\end{itemize}

For consistency, any prescription in the non relativistic approximation should be derived from the relativistic theory as the limiting case $\vert \boldsymbol{\beta} \vert <<1$, where $\boldsymbol{\beta}$ is the particle "velocity" ($c$=1). That is why the positron Schr\"odinger-Pauli equation was obtained in \cite{L1} from the positron Dirac equation, making the substitution $\psi_{C}=e^{i m t}\overset{\sim}{\psi_{C}}$ and taking, then, the non-relativistic limit of the resulting equation.

It was found thus a natural definition for the charge conjugation matrix in the non-relativistic limit,
\begin{equation}
\begin{pmatrix}
\overset{\sim}{\psi_1}\cr\ \overset{\sim}{\psi_2} \cr
\end{pmatrix}
\overset{C}{\longrightarrow}
\begin{pmatrix}
-\overset{\sim}{\psi_2^{*}}\cr\ \overset{\sim}{\psi_1^{*}} \cr
\end{pmatrix}
:=C_{nr}\begin{pmatrix}
\overset{\sim}{\psi_1}\cr\ \overset{\sim}{\psi_2} \cr
\end{pmatrix},
\end{equation}
where:

\begin{equation}
C_{nr}=K
\begin{pmatrix}
0 & -1 \cr 1 & 0 \cr
\end{pmatrix},
\label{matrizCnr}
\end{equation}
with $K$, the complex conjugation operation.

Taking into account the works \cite{L1, L2}, and also some previous studies about the non-relativistic limit in quantum field theory, like the non-relativistic limit of $\lambda {\Phi} ^4$ theory \cite{limite}, we are going to study the non-relativistic limit of charge conjugation, getting a little closer to reality, that is, in the case of the quantum Dirac field coupled to an external electromagnetic field.

\section{On the non-relativistic limit of C in QED}

\hspace{0.5cm} From the \textit{QED} lagrangian density:
\begin{equation}
\textit{L}_{QED}=\bar{\psi}(x)(i\gamma^{\mu}\partial_{\mu}-q \gamma^{\mu}A_{\mu}(x)-m)\psi(x)-\dfrac{1}{4}F_{\mu\nu}(x)F^{\mu\nu}(x);
\label{lagQEDcompleto}
\end{equation}
the equations for the coupled Dirac and electromagnetic fields are deduced (\cite{BandD}, p. 84):

\begin{equation}
(i\gamma^{\mu}\partial_{\mu}-q \gamma^{\mu}\hat{A_{\mu}}(x)-m)\hat{\psi}(x)=0,
\label{ecQED1}
\end{equation}
\begin{equation}
\frac{\partial \hat{F}^{\mu\nu}(x)}{\partial x^{\nu}}=q \hat{\bar{\psi}} \gamma^{\mu}\hat{\psi}(x).
\label{ecQED2}
\end{equation}

In equations (\ref{ecQED1}) and (\ref{ecQED2}), $\hat{A_{\mu}}(x)$ is not considered external and is part of the dynamics of the system. If we make in these equations $\hat{\psi}(x) \longrightarrow \hat{\psi}_{C}(x)$ (equivalent here to change $q$ to $-q$), then changing $\hat{A_{\mu}}(x)\longrightarrow -\hat{A_{\mu}}(x)$ the charge conjugation operation is completed, showing the C invariance of quantum electrodynamics.

If we want now to take the non-relativistic limit of full \textit{QED}, we must establish this limit both for $\hat{\psi}(x)$ and $\hat{A_{\mu}}(x)$. The problem is that it is not possible to establish a non-relativistic limit for $\hat{A_{\mu}}(x)$ because the photon mass is zero.

It is discarded from here that, when we consider electrons, positrons and photons together, makes sense to talk about the charge conjugation operation in the non-relativistic limit. As we can not establish the non-relativistic limit for \textit{QED}, from this point on, we will forget the equation (\ref{ecQED2}) and will assume that the electromagnetic field in (\ref{ecQED1}) is now a classical external object, with perfectly determined value
 $\hat{A_{\mu}}(x)=A_{\mu}$. That is, we are going to study the charge conjugation operation in the context of the Dirac field coupled to a classical external electromagnetic field.

\section{The case of an external electromagnetic field}

\hspace{0.5cm} What was determined in \cite{L1} and \cite{L2} was the expression for the $C$ matrix in the non-relativistic limit of the Dirac equation coupled to an external electromagnetic field. It is also necessary to find now the $\hat{C}$ operator, which represents the charge conjugation operation in the Hilbert space and it is related to the $C$ matrix through $C\hat{\bar{\psi}}^T(x)={\hat C}^{\dag}\hat{\psi}(x){\hat C}$, in the context of the non-relativistic limit of the Dirac field coupled too to a classical external electromagnetic field.

The creation and annihilation operators in terms of the field operators are given by:

\begin{equation}
\hat{b}(\bold{p},r)=\int \dfrac{d^{3}x}{(2 \pi) ^{3/2}}\sqrt{\dfrac{m}{E_{p}}}\overset{-}{u}(\bold{p},r)\hat{\psi}(x)e^{ip\cdot x},
\label{b}
\end{equation}

\begin{equation}
\hat{d}^{\dagger}(\bold{p},r)=\int \dfrac{d^{3}x}{(2 \pi) ^{3/2}}\sqrt{\dfrac{m}{E_{p}}}\overset{-}{v}(\bold{p},r)\hat{\psi}(x)e^{-ip\cdot x},
\label{d dagger}
\end{equation}

\begin{equation}
\hat{b}^{\dagger}(\bold{p},r)=\int \dfrac{d^{3}x}{(2 \pi) ^{3/2}}\sqrt{\dfrac{m}{E_{p}}}\hat{\overset{-}{\psi}}(x)u(\bold{p},r)e^{-ip\cdot x},
\label{b dagger}
\end{equation}

\begin{equation}
\hat{d}(\bold{p},r)=\int \dfrac{d^{3}x}{(2 \pi) ^{3/2}}\sqrt{\dfrac{m}{E_{p}}}\hat{\overset{-}{\psi}}(x)v(\bold{p},r)e^{ip\cdot x}.
\label{d}
\end{equation}

At low speeds $(\frac{\vert \bold{p} \vert}{c}\longrightarrow 0)$ the creation and annihilation operators ($\hat{b}^{\dagger}, \hat{d}^{\dagger}, \hat{b}, \hat{d}$) will continue creating or annihilating particles and antiparticles. Thus, in the non-relativistic limit, the form of the equations (\ref{b}), (\ref{d dagger}), (\ref{b dagger}) and (\ref{d}) will be the same; with the differences that in this case the term $\sqrt{\dfrac{m}{E_{p}}}$ will be equal to  $1$, $\hat{\psi}$ and $\hat{\overset{-}{\psi}}$ will be solutions of the Schr\"odinger-Pauli equations and the spinors will be reduced to:
\begin{eqnarray}
u(\bold{p},1)&=&\begin{pmatrix}
1 \cr 0 \cr 
\end{pmatrix},\quad u(\bold{p},2)=\begin{pmatrix}
0 \cr 1 \cr 
\end{pmatrix},\nonumber\\
\overset{-}{u}(\bold{p},1)&=&\begin{pmatrix}
1, \; 0 \end{pmatrix},\quad
\overset{-}{u}(\bold{p},2)=\begin{pmatrix}
0, \; 1
\end{pmatrix},
\end{eqnarray}
and
\begin{eqnarray}
v(\bold{p},1)&=&\begin{pmatrix}
1 \cr 0 \cr
\end{pmatrix},\quad v(\bold{p},2)=\begin{pmatrix}
0 \cr 1 \cr
\end{pmatrix},\nonumber\\
\overset{-}{v}(\bold{p},1)&=&\begin{pmatrix}
-1, \; 0 \end{pmatrix},\;
\overset{-}{v}(\bold{p},2)=\begin{pmatrix}
0, \; -1
\end{pmatrix}.
\end{eqnarray}

The $\hat{C}$ operator corresponding to the Dirac field, as a function of the creation and annihilation operators is given by \cite{Greiner}:

\begin{eqnarray}
\hat{C_{0}}=exp(i\frac{\pi}{2}\int d^{3}p \sum_{r=1}^{2}(\hat{d}^{\dagger}(\bold{p},r)\hat{b}(\bold{p},r)+\hat{b}^{\dagger}(\bold{p},r)\hat{d}(\bold{p},r)\nonumber\\-\hat{b}^{\dagger}(\bold{p},r)\hat{b}(\bold{p},r)-\hat{d}^{\dagger}(\bold{p},r)\hat{d}(\bold{p},r)));
\label{C operador}
\end{eqnarray}
while in the presence of interactions, the charge conjugation operator at time $t$ is given by (\cite{BandD}, p. 111):

\begin{equation}
\hat{C}(t)=e^{i\hat{H_{T}}t}\hat{C_{0}}e^{-i\hat{H_{T}}t},
\label{Cdt}
\end{equation}
where $\hat{H_{T}}$ is the total hamiltonian (Dirac field coupled to the external electromagnetic field) and $\hat{C_{0}}$ satisfies (\ref{C operador}), keeping the operators $\hat{b}^{\dagger}, \hat{d}^{\dagger}, \hat{b}, \hat{d}$ in the Heisenberg picture.

By taking the non-relativistic limit, the form of the operator $\hat{C}(t)$, given by (\ref{Cdt}), does not change; but then, the operators $\hat{b}^{\dagger}, \hat{d}^{\dagger}, \hat{b}, \hat{d}$ will be those for low speeds, and $\hat{H_{T}}$ will be the Schr\"odinger-Pauli hamiltonian.

Summarizing, in a relativistic context the charge conjugation operation is equivalent to change $\hat{\psi}(x)\longrightarrow\hat{\psi}_{C}(x)$ and  $\hat{A}_{\mu}(x)\longrightarrow-\hat{A}_{\mu}(x)$, both for $\hat{A}_{\mu}(x)$ dynamics and for $\hat{A}_{\mu}(x)={A}_{\mu}$ classical external. This is equivalent to say that, in a relativistic context, the electron move in a field $\hat{A}_{\mu}(x)$ like the positron does in a field $-\hat{A}_{\mu}(x)$.

In the non-relativistic limit, the charge conjugation operation only makes sense for ${A}_{\mu}$ classical external and it is reduced only to change $\hat{\psi}(x)\longrightarrow\hat{\psi}_{C}(x)$. In this last case, to make also the change $A_{\mu}\longrightarrow-A_{\mu}$ do not turn the Schr\"odinger-Pauli equation for the positron into the Schr\"odinger-Pauli equation for the electron. 

\section{Conclusions}

\hspace{0.5cm} Even if it is not possible to establish the charge conjugation operation in the non-relativistic limit of full quantum electrodynamics, as is not possible to take this limit for the electromagnetic field because the photon mass is zero, we can talk about the non-relativistic limit of charge conjugation for the case of the Dirac field in the presence of a classical external electromagnetic field and give the expressions, for both the $C$ matrix and the $\hat{C}$ operator, in this case.

Regarding whether the process of renormalization could modify the above analysis, we should take into account that we are not studying the elements of the dispersion matrix in each approximation of perturbation theory. Instead, we are talking about the non-relativistic limit of the hamiltonian, with which is possible to calculate the dispersion matrix. Thus, renormalization does not affect the process of taking the limit.

\section{Acknowledgment}
This work was partially support by the project PAPIIT IN 118609-2, DGAPA-UNAM, México. B. Carballo Pérez also acknowledges financial support from CONACyT, México.


\end{document}